\documentclass[preprint,aps]{revtex4}
\usepackage{graphicx}
\begin{document}

\title{Application prospects of MgB$_2$ in view of its basic properties}

\author{Michael Eisterer and Harald W. Weber}

\address{Atominstitut der \"Osterreichischen Universit\"aten, Vienna University of Technology, 1020
Vienna, Austria}

\begin{abstract} Seven years after the discovery of
superconductivity in magnesium diboride, the fundamental
superconducting properties of this compound are well known and the
peculiar current transport in polycrystalline materials is
essentially understood. Based on this knowledge the ultimate
performance of wires or tapes at high magnetic fields will be
predicted and compared to state-of-the-art materials and to other
superconductors. The key parameter for high field applications is
the upper critical field, which can be strongly enhanced by
impurity scattering. This fundamental property might be further
optimized in bulk materials, since higher values were reported for
thin films. The MgB$_2$ grains are usually very small, if prepared
by the in-situ technique. The resulting high density of grain
boundaries leads to strong pinning, close to the theoretical
limit. On the other hand, the connectivity between the grains is
still rather poor and strongly reduces the achievable critical
currents, thus leaving room for further improvements.

\end{abstract}

\maketitle

\section{Introduction}

The critical current density $J_\mathrm{c}$ is the most important
quantity for power applications. Its magnitude is given by the
pinning strength and its field dependence is determined by the
density and morphology of pins and by the upper critical field. At
low magnetic fields, $J_\mathrm{c}$ can be around 15--20\% of the
depairing current density $J_\mathrm{d}$ in a type II
superconductor with optimized pinning  \cite{Arc05}. This limit is
obtained from the balance of the maximum force a single pin can
exert on a flux line and the Lorentz force acting on this ideally
pinned vortex. As long as the density of such strong pins is
significantly higher than the density of vortices, $J_\mathrm{c}$
can meet this theoretical limit. At higher magnetic fields, not
all flux lines are perfectly pinned anymore and $J_\mathrm{c}$
starts to decrease. This is illustrated in Fig.~\ref{Figisoaniso}
(open squares). The low field $J_\mathrm{c}$ is given by the
pinning force of an individual pin or, more precisely, by the sum
of the forces of all pins acting on the same flux line. The field
$B_\mathrm{sv}$, where the plateau ends, is determined by the pin
density. At higher fields, the field dependence is given by the
morphology of the defects \cite{Dew74} and by the upper critical
field $B_\mathrm{c2}$. For the sake of simplicity, the critical
currents are assumed to reach zero at $B_\mathrm{c2}$, since
thermal fluctuations are unimportant in MgB$_2$ \cite{Kos05}.

The field dependence of the critical current densities plotted in
Fig.~\ref{Figisoaniso} is predicted for grain boundary pinning
\cite{Dew74} which is most important in MgB$_2$
\cite{Mar07,Eis07rev} and will be considered in the following.
$J_\mathrm{c}$ is given by
\begin{equation}
J_\mathrm{c}(B)=A_\mathrm{con}\eta_\mathrm{pin}J_\mathrm{d}\frac{(1-\frac{B}{B_\mathrm{c2}})^2}{\sqrt{B}}
\label{Jcmodel}
\end{equation}
for fields above $B_\mathrm{sv}$. $A_\mathrm{con}$ is one in the
ideal case and describes the suppression of the current density
due to a reduction of the effective cross section resulting from
porosity, secondary phases or badly connected grains \cite{Row03}.
$\eta_\mathrm{pin}$ characterizes the pinning efficiency.
$A_\mathrm{con}$ and $\eta_\mathrm{pin}$ are not intrinsic to a
certain material but given by the actual pinning centers or
microstructure. With optimized values for these extrinsic
parameter, $J_\mathrm{c}(B)$ is entirely determined by the two
intrinsic parameters $J_\mathrm{d}$ and $B_\mathrm{c2}$. The
depairing current density scales the y-axis in
Fig.~\ref{Figisoaniso}, the upper critical field scales the
x-axis. These fundamental parameters of MgB$_2$ are compared to
those of other technologically important superconductors in
Table~\ref{tabprop}. All values refer to zero temperature. The
depairing current density in MgB$_2$ is much higher than in NbTi
and even comparable to that in high temperature superconductors.
Only Nb$_3$Sn has a significantly larger depairing current
density. Because of the two band nature of superconductivity in
MgB$_2$ and the suppression of the $\pi$ gap at high magnetic
fields, $J_\mathrm{d}$ slightly decreases with magnetic field (to
about $1.3\times 10^{12}$\,A\,m${^2}$ \cite{Eis07rev}).

The upper critical field at 0\,K is rather low in clean MgB$_2$
($\sim$14\,T), comparable to NbTi. Note that the given value
refers to the apparent $B_\mathrm{c2}$ in polycrystalline MgB$_2$,
which corresponds to $B_\mathrm{c2}^{ab}$ (field parallel to the
boron planes). Disorder enhances the upper critical field
significantly, as discussed in Section~\ref{SecDis}, and values of
up to 74 T were reported \cite{Bra05}.

These two fundamental parameters ($J_\mathrm{d}$, $B_\mathrm{c2}$)
indicate that MgB$_2$ is a promising high field conductor.
Unfortunately, the anisotropic magnetic properties of magnesium
diboride complicate the current transport and reduce the critical
currents at high magnetic fields significantly.

\section{Upper Critical Field Anisotropy}

\begin{figure} \centering
\includegraphics[width=3.2in,clip]{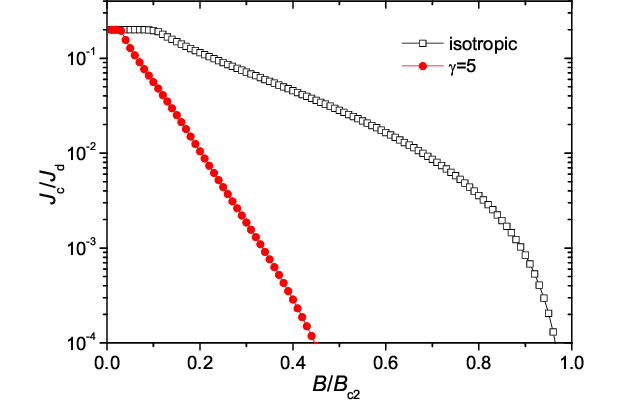} \caption{Influence
of the upper critical field anisotropy on the critical current
densities. The field and current densities are normalized by the
upper critical field and the depairing current density,
respectively. Calculations are based on grain boundary pinning.}
\label{Figisoaniso}
\end{figure}

The upper critical field anisotropy, $\gamma$, changes the field
dependence of the critical current density \cite{Eis03} and of the
corresponding volume pinning force \cite{Eis08} drastically. This
is illustrated in Fig.~\ref{Figisoaniso}. The solid circles
represent $J_\mathrm{c}$ in a polycrystalline material with an
anisotropy factor of 5, which is representative for clean MgB$_2$.
The field, $B_{\rho=0}$, where the critical currents reach zero,
decreases by a factor of $((\gamma^2-1)p_\mathrm{c}^2+1)^{-1/2}$
\cite{Eis05} compared to the corresponding isotropic material and
the field dependence of $J_\mathrm{c}$ is enhanced. The
percolation threshold $p_\mathrm{c}$ denotes the minimum fraction
of superconducting material for a continuous current path and is
expected to be 0.2 to 0.3 \cite{Eis03}. In the following
$p_\mathrm{c}=0.25$ is assumed, leading to a reduction of
$B_{\rho=0}$ to $0.63B_\mathrm{c2}$ (for $\gamma=5$). All
calculations are based on the model proposed in \cite{Eis03}, but
with the anisotropic scaling approach \cite{Bla92}, as described
in \cite{Eis08}.

\begin{table}
\renewcommand{\arraystretch}{1.3}
\caption{Depairing current densities and upper critical fields of
various superconductors at zero temperature} \label{tabprop}
\centering
\begin{tabular}{l c c c c}
\hline & MgB$_2$ & NbTi & Nb$_3$Sn & HTS \\ \hline
$J_\mathrm{d}$($10^{12}$\,A\,m$^{-2}$)&2&0.4&7&3\\
$B_\mathrm{c2}$(T)&14-70 ($H\|ab$) &15&30&150 ($H\|c$) \\ \hline
\end{tabular}
\end{table}

The anisotropy is the third intrinsic parameter which influences
the critical currents in MgB$_2$.

\section{Disorder improves the high field properties \label{SecDis}}

The low upper critical field and the high anisotropy of clean
MgB$_2$ restrict high currents to low magnetic fields
(cf.~Fig.~\ref{Figisoaniso}). Impurity scattering helps to solve
this problem by enhancing the upper critical field and reducing
its anisotropy at the expense of a slight decrease in
$T_\mathrm{c}$. This reduction of the transition temperature is
caused by interband scattering between the $\sigma$- and the
$\pi$-bands and by intraband scattering within the $\sigma$-bands.
Intraband scattering can reduce $T_\mathrm{c}$ only in anisotropic
superconductors or, indirectly,  via a reduction of the density of
states (DOS) at the Fermi level, as observed in MgB$_2$
\cite{Put07}. Since intraband scattering in the $\sigma$-bands is
also responsible for the increase in the upper critical field, a
decrease in $T_\mathrm{c}$ is inherent in the enhancement of
$B_\mathrm{c2}$ by impurity scattering. The additional amount of
interband scattering potentially depends on the actual defect
structure and leads to sample to sample variations. The same holds
for charge doping (e.g. by carbon or aluminium), which also
reduces the DOS \cite{Pen02,Umm05b,Kor05}. Nevertheless, a
remarkable similarity in resistivity, transition temperature and
upper critical field was found in samples with a totally different
defect structure (as grown defects, SiC doping, neutron induced
defects) \cite{Eis07}. Thus, the transition temperature turns out
to be a useful disorder parameter.

The beneficial effect of the decrease of the mean free path of the
charge carriers has to compete with the decrease of $T_\mathrm{c}$
leading to a maximum of $B_\mathrm{c2}$ as a function of the
transition temperature. $B_\mathrm{c2}$ in most samples with a
transition temperature above 33\,K ($B_\mathrm{c2}<\sim$ 40\,T)
reasonably follows the relation \cite{Eis07rev}
\begin{equation}
B_\mathrm{c2}(T_\mathrm{c})=13.8(t_\mathrm{c}^2+16.7t_\mathrm{c}(1-t_\mathrm{c}))\,\mathrm{T},
\label{Bc2Tc}
\end{equation}
with
$t_\mathrm{c}:=\frac{T_\mathrm{c}}{T_\mathrm{c}^\mathrm{clean}}$
and $T_\mathrm{c}^\mathrm{clean}=39.43$\,K. $B_\mathrm{c2}$ of
samples with a smaller transition temperature normally deviates
from this relation, as demonstrated by irradiation
\cite{Gan05b,Tar06,Kru07} or doping \cite{Wil04,Ang05,Kru06}
experiments. Only carbon doped fibers \cite{Fer05} follow
(\ref{Bc2Tc}) at lower transition temperatures, with
$B_\mathrm{c2}=55$\,T at $T_\mathrm{c}=28$\,K, which is close to
the predicted maximum of about 60\,T at $T_\mathrm{c}\sim 21$\,K.
The doped fibers and results obtained on thin films, where upper
critical fields of up to 74\,T were found \cite{Bra05}, suggest
that the highest reported values in bulk samples (around 40\,T)
are not a fundamental limit and that the deviations from
(\ref{Bc2Tc}) at low transition temperatures are caused by an
additional not inherent effect (e.g. strong interband scattering).
However, the validity of (\ref{Bc2Tc}) down to low transition
temperatures will be assumed in the following, which is
optimistic, since the maximum $B_\mathrm{c2}$ at around 21\,K was
not yet achieved in bulk samples. On the other hand, many reports
claim slightly higher upper critical fields at a certain
transition temperature ($>$\,33\,K) than predicted by
(\ref{Bc2Tc}), which reflects the \emph{average} behavior of data
extracted from literature \cite{Eis07rev}. Ideal intraband
scattering centers inducing only a minimum fraction of interband
scattering could lead to higher values of the upper critical field
at a given transition temperature than predicted by (\ref{Bc2Tc}).

A further benefit of impurity scattering is the reduction of the
upper critical field anisotropy \cite{Kru07} which is also
strongly correlated with the reduction of $T_\mathrm{c}$
\cite{Kru07,Eis07rev}:
\begin{equation}
\gamma(T_\mathrm{c})=\frac{t_\mathrm{c}^2+16.7t_\mathrm{c}(1-t_\mathrm{c})}{3.88-3.724t_\mathrm{c}}.
\label{gammaTc}
\end{equation}

\begin{figure} \centering
\includegraphics[width=3.2in,clip]{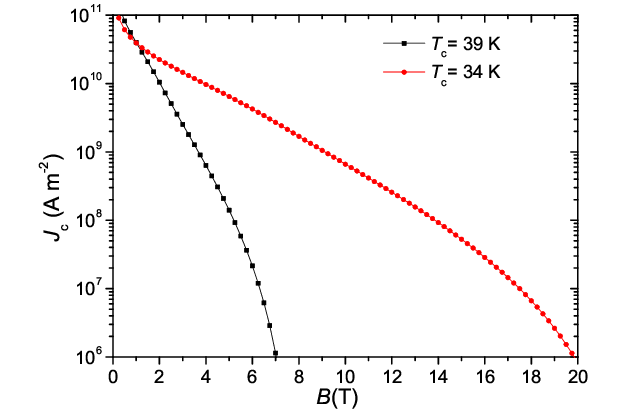} \caption{Influence
of disorder on the critical current densities. $T_\mathrm{c}$
serves as disorder parameter.} \label{FigJcTc}
\end{figure}

The depairing current density is expected to decrease by the
introduction of disorder, since $J_\mathrm{d}\propto
B_\mathrm{c}^2 \xi \propto T_\mathrm{c}^2/\sqrt{B_\mathrm{c2}}$.
Both, the reduction of $T_\mathrm{c}$ and the enhancement of
$B_\mathrm{c2}$, decrease $J_\mathrm{d}$. The estimate
$T_\mathrm{c} \propto B_\mathrm{c}$ is based on the
proportionality between the thermodynamic critical field,
$B_\mathrm{c}$, and the energy gap, which was shown to scale with
$T_\mathrm{c}$ in the $\sigma$-band \cite{Gon05,Zam05,Put06}. With
this dependence of $J_\mathrm{d}$ on $T_\mathrm{c}$ and
$B_\mathrm{c2}$ and with the dependencies of $B_\mathrm{c2}$ and
$\gamma$ according to (\ref{Bc2Tc}) and (\ref{gammaTc}), the
critical current density (at fixed field and temperature) is a
unique function of the transition temperature for a given
$A_\mathrm{con}\eta_\mathrm{pin}$. Figure~\ref{FigJcTc} compares
the field dependence of the critical currents in clean
($T_\mathrm{c}=39$\,K) and disordered ($T_\mathrm{c}=34$\,K)
MgB$_2$.

\section{Optimization of the material properties}

\begin{figure} \centering
\includegraphics[width=3.2in,clip]{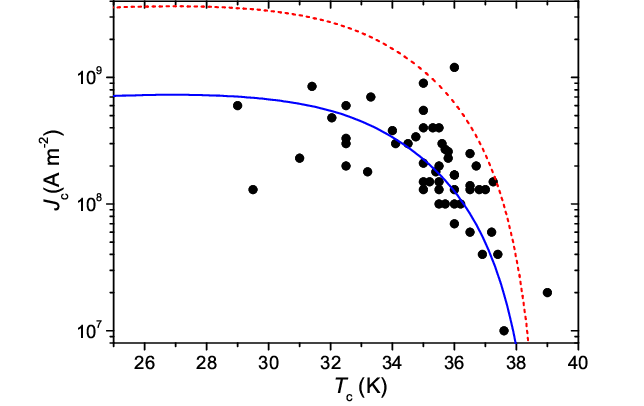} \caption{Critical
current densities at 4.2 K and 8 T as a function of the transition
temperature. Experimental data are shown for comparison (solid
circles). They were extracted from literature in \cite{Eis07rev}
and a view more recent data were added in this plot
\cite{Her07,Her07b,His07,Kim07b,Kov07b,Sen07,Wan07b,Fuj08,Hur08,Sen08,Shc08,Zha08}.
\label{FigJc8T}}
\end{figure}

Since the improvements in $B_\mathrm{c2}$ and $\gamma$ induced by
impurity scattering have to compete with the reduction of
$J_\mathrm{d}$ and $T_\mathrm{c}$, an optimal amount of disorder
maximizes the critical current density at a certain field and
temperature. This is illustrated in Figure~\ref{FigJc8T} for 8\,T
and 4.2\,K .

The only free parameter left is the product of $A_\mathrm{con}$
and $\eta_\mathrm{pin}$, which is temperature independent and can
be chosen to fit the experimental data (solid line in
Fig.~\ref{FigJc8T}). The overall agreement between the
experimental data and the predicted behavior is quite good, given
the fact that $A_\mathrm{con}$ and $\eta_\mathrm{pin}$ are
expected to vary from sample to sample. The main increase of the
high field $J_\mathrm{c}$ was achieved by the introduction of
scattering centers in the last few years, which decreased
$T_\mathrm{c}$ to about 32--35\,K. A further decrease of
$T_\mathrm{c}$ seems to be less beneficial, although the predicted
maximum of $J_\mathrm{c}$ is at around 27\,K.

It is interesting to compare the experimental data to the expected
maximum of $A_\mathrm{con}\eta_\mathrm{pin}$.  $A_\mathrm{con}$ is
one in the ideal case, but it is difficult to estimate the
theoretical limit of $\eta_\mathrm{pin}$. As pointed out in the
introduction, the maximum low field $J_\mathrm{c}$ is expected to
be around 0.2$J_\mathrm{d}$. Due to the divergence of
$J_\mathrm{c}$ in (\ref{Jcmodel}) for $B\rightarrow0$,
$A_\mathrm{con}\eta_\mathrm{pin}$ cannot be determined from the
self field $J_\mathrm{c}$ alone. The field, where the low field
plateau ends, is crucial and depends on the density of strong
pins. It should be pointed out that the low field behaviour of the
critical currents is difficult to determine, since the self field
of the currents becomes quite large at 0.2$J_\mathrm{d}\approx
4\times 10^{11}$ A\,m$^{-2}$. Approximately 1\,T can be estimated
as the maximum self field for a 400\,nm thick film (only weakly
dependent on its lateral dimensions) and 200\,T (!) for a cube of
1\,mm$^3$. The latter is of course totally unrealistic, since the
self field strongly reduces the currents leading in turn
immediately to much smaller self fields. Nevertheless, the self
field remains quite large in bulk samples and impedes the
assessment of $J_\mathrm{c}$ at low fields. Self field current
densities of around $4\times 10^{11}$ A\,m$^{-2}$ were reported in
thin films \cite{Kim01} indicating the presence of optimal pinning
centers. $J_\mathrm{c}$ is usually about one order of magnitude
lower at 1\,T \cite{Eis07rev} which might result from a
comparatively low density of such pins. The mean distance between
two pins should be significantly smaller than the lattice
parameter of the flux line lattice which is about 44\,nm at 1\,T.
However, from $J_\mathrm{c}(1\,\mathrm{T},
4.2\,\mathrm{K})=4\times 10^{10}$ A\,m$^{-2}$ one can estimate
$A_\mathrm{con}\eta_\mathrm{pin}$ to be about 0.018\,T$^{0.5}$.
The dotted line in Fig.~\ref{FigJc8T} is based on this value,
which is about 5 times larger than the average experimental one in
polycrystalline materials (solid line). It is in principle not
possible to distinguish whether pinning is less efficient in such
materials or whether the connectivity is reduced in comparison to
thin films. The typical low densities of the filaments and the
presence of secondary phases favor the latter \cite{Eis07rev},
which is supported by electron microscopy \cite{Bir08}. However,
$A_\mathrm{con}\eta_\mathrm{pin}=0.018\,$T$^{0.5}$ is used in the
following, which is not a theoretical limit, but corresponds to
the highest reported critical current densities in well connected
films. Nanometer sized grains were found not only in these ``high
$J_\mathrm{c}$'' films, but also in bulks, wires or tapes. It is
interesting to note that two experimental points
\cite{Hur08,Giu03}, lie \emph{above} this ``optimal'' performance.
Both samples were prepared by an internal magnesium diffusion
process. (The data point at 39\,K extracted from Ref.~\cite{Giu03}
was slightly extrapolated from lower field values. $T_\mathrm{c}$
is given only approximately ($\approx 36$\,K) in
Ref.~\cite{Hur08}.) This gives hope that this prediction is still
too pessimistic, although it might be simply a consequence of a
slight overestimation of $T_\mathrm{c}$ because of material
inhomogeneity. In this case the grains or regions of the sample
which determine the onset of the superconducting transition are
not responsible for the high field behavior at low temperatures
\cite{Eis07rev}.

\begin{figure} \centering
\includegraphics[width=3.2in,clip]{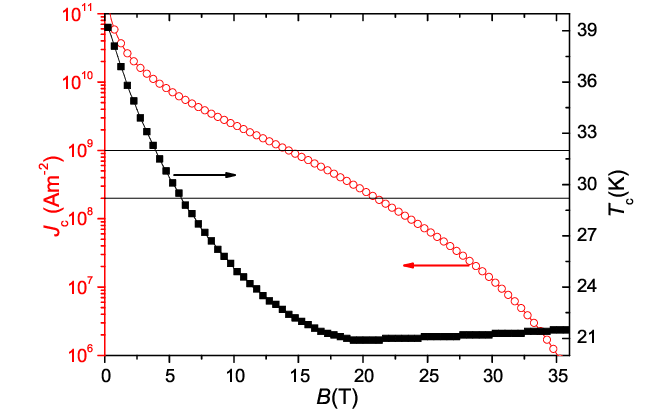} \caption{Optimum
critical current densities at 4.2\,K (open circles) and
corresponding transition temperatures (solid squares). Horizontal
lines indicate typical application criteria. \label{FigJcoptTc}}
\end{figure}

The optimal amount of scattering centers depends on the operating
conditions. The potentially highest $J_\mathrm{c}$ at each field
was calculated and is presented in Fig.~\ref{FigJcoptTc} (open
circles). A clean material with optimum $T_\mathrm{c}$ is
favorable at low fields. With increasing field, disorder should
increase and the optimum $T_\mathrm{c}$ is predicted to be about
21\,K for magnetic fields above about 20\,T. This implicitly
assumes a $B_\mathrm{c2}$ of 47\,T at 4.2 K ((\ref{Bc2Tc}) and
(\ref{Bc2T})), which was only realized in fibers \cite{Fer05} or
films \cite{Bra05} so far (although at higher transition
temperatures). If $B_\mathrm{c2}$ cannot be improved beyond 40\,T,
the optimum high field $J_\mathrm{c}$ is expected to be similar to
that of the ``34\,K sample'' in Fig.~\ref{FigJcTc}.
($A_\mathrm{con}\eta_\mathrm{pin}=0.018\,$T$^{0.5}$ was assumed
also in these calculations.)

Typical magnet applications require a current density
$J_\mathrm{c}$ of about $10^{9}$\,A\,m$^{-2}$, for large magnets
(e.g. in fusion power plants) $2\times 10^{8}$\,A\,m$^{-2}$ might
be acceptable.  At 4.2\,K, these criteria are met at 14\,T and
21\,T, respectively.

\begin{figure} \centering
\includegraphics[width=3.2in,clip]{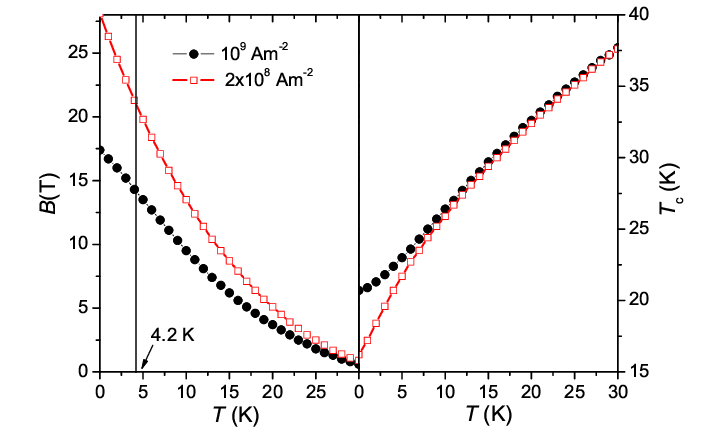} \caption{Left panel: Fields and
temperatures where the optimum $J_\mathrm{c}$ is
$10^{9}$\,A\,m$^{-2}$ (solid circles) and $2\times
10^{8}$\,A\,m$^{-2}$ (open squares). Right panel: Corresponding
transition temperature. \label{FigAppl}}
\end{figure}

MgB$_2$ could also be used at higher temperatures, thus the
temperature dependence of the ``application fields'' was
calculated. The results are shown in Fig.~\ref{FigAppl}. The
depairing current density was assumed to decrease with temperature
as $J_\mathrm{d}(T)\propto B_\mathrm{c}(T)^2 \xi(T) \propto
(1-(T/T_\mathrm{c})^2)^2/\sqrt{B_\mathrm{c2}(T)}$. The upper
critical field was modeled by
\begin{equation}
B_\mathrm{c2}(T)=B_\mathrm{c2}(0)(1-T/(T_\mathrm{c}-4\,\mathrm{K})).
\label{Bc2T}
\end{equation} A linear dependence of $B_\mathrm{c2}$ on
temperature is often observed in MgB$_2$ at intermediate
temperatures. This linear behavior does not extrapolate to zero at
$T_\mathrm{c}$, but at a slightly smaller temperature, therefore,
4\,K are subtracted from $T_\mathrm{c}$ in the denominator in
(\ref{Bc2T}). The usually observed tail in $B_\mathrm{c2}(T)$ near
$T_\mathrm{c}$ is neglected, since the temperature range near
$T_\mathrm{c}$ is unimportant for applications. Note that the
experimental data of $B_\mathrm{c2}(0)$, on which (\ref{Bc2Tc}) is
based, were often obtained from a linear extrapolation of
$B_\mathrm{c2}(T)$ at intermediate temperatures, which makes the
present estimation of $B_\mathrm{c2}(T)$ consistent with these
data.

The critical current density is above $10^{9}$\,A\,m$^{-2}$ up to
about 3.7\,T at 20\,K and above $2\times 10^{8}$\,A\,m$^{-2}$ up
to about 5 T (Fig.~\ref{FigAppl}). These values should not be
taken too literally, because of the numerous assumptions.

The optimum transition temperature increases with the application
temperature and depends only weakly on the application criterion.
Only at low temperatures significant differences between the
$10^{9}$\,A\,m$^{-2}$ and the $2\times 10^{8}$\,A\,m$^{-2}$ curves
are found (right panel in Fig.~\ref{FigAppl}). The optimum
transition temperature at liquid helium temperature is still much
lower than in state-of-the art materials (if (\ref{Bc2Tc}) could
be realized). At 20\,K the optimum $T_\mathrm{c}$ is predicted to
be around 32.5\,K which is comparable to today's materials. Thus,
a further increase of impurity scattering is not expected to
improve the properties at 20\,K. This can be realized only by
improving the connectivity or pinning.

\begin{figure} \centering
\includegraphics[width=3.2in,clip]{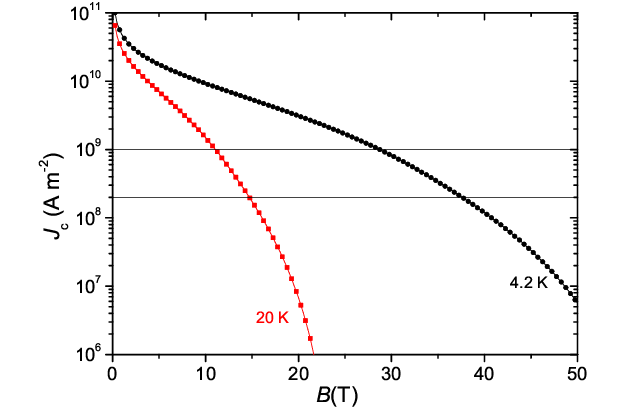}
\caption{A more optimistic scenario: Calculations are based on the
best thin film performance reported so far. \label{Filmtheo}}
\end{figure}

Last we consider an even more optimistic scenario based on the
best demonstrated thin film performance. An upper critical field
of around 63\,T at 4.2\,K with an anisotropy of about 1.85 was
found in a thin film with $T_\mathrm{c}=35$\,K \cite{Bra05}.
$B_\mathrm{c2}$ of the same film was 27\,T at 20\,K with an
anisotropy of about 2.2. With these parameters and
$A_\mathrm{con}\eta_\mathrm{pin}$ as above (demonstrated at
\emph{other} films), $J_\mathrm{c}(B)$ was calculated, assuming
the same dependence of $J_\mathrm{d}$ on $T$, $T_\mathrm{c}$ and
$B_\mathrm{c2}$. The results are shown in Fig.~\ref{Filmtheo}. The
application field is shifted to around 30\,T at 4.2\,K and to
about 13\,T at 20\,K.

\section{Conclusions}

The optimum high field performance of MgB$_2$ was calculated based
on its intrinsic parameters, namely upper critical field,
depairing current density, anisotropy and transition temperature.
It turns out that the optimal amount of impurity scattering
depends on the operation conditions (field and temperature). The
upper critical field and the anisotropy were extrapolated from
literature data, which only reflect an average behavior.
Therefore, further improvements can be expected by optimization of
the inter-/intraband scattering rates. This is encouraged by
results obtained on thin films. Other promising ways of enhancing
the critical currents are an improvement in the connectivity
between the grains and of the MgB$_2$ density in wires or tapes.
At low temperatures, MgB$_2$ has certainly the potential to beat
the high field performance of NbTi and even to approach that of
Nb$_3$Sn. At 20\,K intraband scattering seems to be nearly
optimized in today's state-of-the-art samples.

If the high upper critical fields observed in dirty thin films
with rather high transition temperatures could be obtained also in
polycrystalline MgB$_2$, the high field performance would
significantly improve.

%\newpage

%\section*{Acknowledgment}



% http://www.ctan.org/tex-archive/macros/latex/contrib/supported/IEEEtran/bibtex
\bibliographystyle{IEEEtran}
%\bibliography{ref}

\end{document}